\documentclass{aa}
\usepackage{graphicx}
\usepackage{natbib}
\usepackage{txfonts}
\begin{document}
   \title{\object{HS\,0702+6043}: A star showing both short-period
     p-mode and long-period g-mode oscillations\thanks{Based on observations
       collected at the Centro 
       Astron\'omico Hispano Alem\'an (CAHA) at Calar Alto,
       operated jointly by the Max-Planck Institut f\"ur
       Astronomie and the Instituto de Astrof\'{\i}sica de Andaluc\'{\i}a
       (CSIC), and at the Bok Telescope on Kitt Peak, operated
       by Steward Observatory (University of Arizona).}}
   
   \subtitle{}

   \author{S. Schuh\inst{1} 
          \and
          J. Huber\inst{1}
          \and
          S. Dreizler\inst{1}
	  \and
          U. Heber\inst{2}
	  \and
          S.J. O'Toole\inst{2}
	  \and
	  E.M. Green\inst{3}
	  \and
	  G. Fontaine\inst{4}
          }

   \offprints{S.\ Schuh}

   \institute{Institut f\"ur Astrophysik, Universit\"at G\"ottingen,
              Friedrich-Hund-Platz~1, D--37077 G\"ottingen, Germany\\ 
	      \email{schuh@astro.physik.uni-goettingen.de}
         \and Dr. Remeis-Sternwarte Bamberg, Universit\"at
              Erlangen-N\"urnberg, Sternwartstra\ss e 7, D--96049
              Bamberg, Germany
         \and Steward Observatory, University of Arizona,
              933~N~Cherry~Ave., Tucson~AZ~85721-0065, USA  
	 \and D{\'e}partement de physique, Universit{\'e} de Montr{\'e}al,
	      CP~6128, Succ centre-ville, Montr{\'e}al, P.Q., Canada, H3C-3J7 
   }
   \date{Received ; accepted}
   \abstract
   {The hot subdwarf B star \object{HS\,0702+6043} is known as a
    large-amplitude, short-period $p$-mode pulsator of the \object{EC~14026}
    type. Its atmospheric parameters place it at the common boundary
    between the empirical instability regions of the \object{EC~14026}
    variables and the typically cooler long-period $g$-mode pulsators of
    the \object{PG\,1716} kind.} 
   {We analyse and interpret the photometric variability of
   \object{HS\,0702+6043} in order to explore its asteroseismological potential.}
   {We report on rapid wide band CCD photometric observations to
   follow up on and confirm the serendipitous discovery of
   multiperiodic long-period luminosity variations with typical time
   scales of $\sim$1\,h in \object{HS\,0702+6043}, in addition to the two previously known
   pulsations at 363\,s and 383\,s. In particular, we isolate a relatively
   low-amplitude ($\sim$4 mmag), long-period (3538$\pm$130\,s) light variation.} 
   {We argue that the most likely origin for this luminosity
    variation is the presence of an excited g-mode pulsation.
   If confirmed, \object{HS\,0702+6043} would constitute a rare
   addition to the very select class of pulsating stars showing
   simultaneously parts of their pressure and gravity mode pulsation
   spectra. The asteroseismological potential of such stars is
   immense, and \object{HS\,0702+6043} thus becomes a target of choice
   for future investigations. While our discovery appears
   consistent with the location of \object{HS\,0702+6043} at the common boundary
   between the two families of pulsating sdB stars, it does challenge
   theory's current description of stability and driving mechanisms in
   pulsating B subdwarfs.}
   {}

\maketitle
\section{Introduction}
\label{sec:introduction}
Subdwarf B stars populate the extreme horizontal branch (EHB) in the
effective temperature range of 22\,000 to 40\,000\,K and have
surface gravity values from  $\log{g}$=5.0 to 6.2 in cgs units.
The masses of these hot, evolved objects should cluster around
0.5~M$_{\sun}$ as suggested by evolution theory \citep{2003MNRAS.341..669H}. 
Indeed, masses between 0.457 and 0.49~M$_{\sun}$
have been derived recently for four \object{EC~14026} stars by asteroseismology
\citep{charp_lapalma:05}.
They are believed to be core helium-burning but
with hydrogen envelopes too thin to sustain H-shell burning.
Standard tracks of stellar evolution do not cross
the EHB region since they do not produce inert hydrogen envelopes, but
the high fraction of binaries among the sdB stars suggests that close
binary evolution may play an important role in their formation.
The relative importance of several proposed feeder channels remains
unclear. 
\par
To learn more about sdB structure and hence their evolutionary
history, asteroseismology is one of the important methodical
approaches, made possible due to short-period oscillations exhibited
by a fraction of sdB stars (called sdBVs). To the group of pulsating
sdB stars known as \object{V631~Hya} variables (more commonly referred to as
\object{EC~14026} variables) which show periods of the order 
of a few minutes, \citet{2003ApJ...583L..31G} have recently added a
longer-period group (tentatively called lpsdBV) for which the
prototype is \object{PG\,1716$+$426}. 
In the short-period sdBVs, a $\kappa$ mechanism drives the low-order
$p$-modes, where the required opacity bump is due to iron accumulated by
diffusion \citep{charpinet:96}. According to \citet{2003ApJ...597..518F}, the same
$\kappa$ mechanism can also drive high-order $g$-modes in the
long-period \mbox{sdBVs} that have periods of the order of about an
hour. 
\par
The m$_B$=15 star \object{HS\,0702$+$6043} was identified as a member of the
\object{EC~14026} group in a search program by \citet[ hereafter DR02]{2002A&A...386..249D}. Its
effective temperature was determined to be 28\,400\,K, its surface gravity
$\log{g}$=5.35. These spectroscopic parameters place it at the cool end of the
\object{EC~14026} instability region. The main pulsation period was found to be
363\,s at a relatively large amplitude of 29\,mmag. A second period of
382\,s has been shown to be present at a much smaller amplitude of
3.8\,mmag. This paper reports on the discovery of additional
variations at a longer time scale of $\sim$1\,h, and
addresses in particular the implications resulting in the case of an
interpretation of the period as a stellar oscillation.
\section{Discovery and follow-up observations}
\label{sec:serendip}
\begin{figure}
  \centering
  \includegraphics[width=0.4\textwidth]{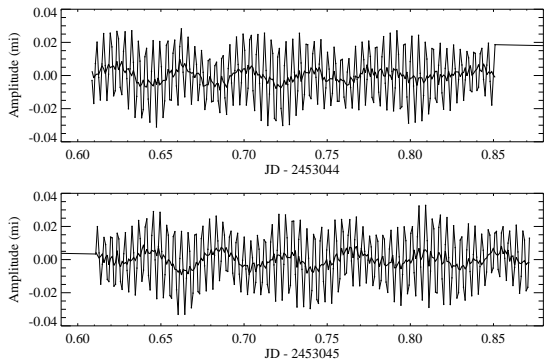}
  \caption{Original data points (dots joined with a line), and
    residuals (thick line) after subtraction of two short-period
    frequencies $f_1$ and $f_2$: The variability pattern of the
    residuals strongly resembles the multiperiod brightness changes
    observed in lpsdBV stars.}
  \label{fig:lc}
\end{figure}
\begin{figure*}
  \centering
  \includegraphics[width=0.8\textwidth]{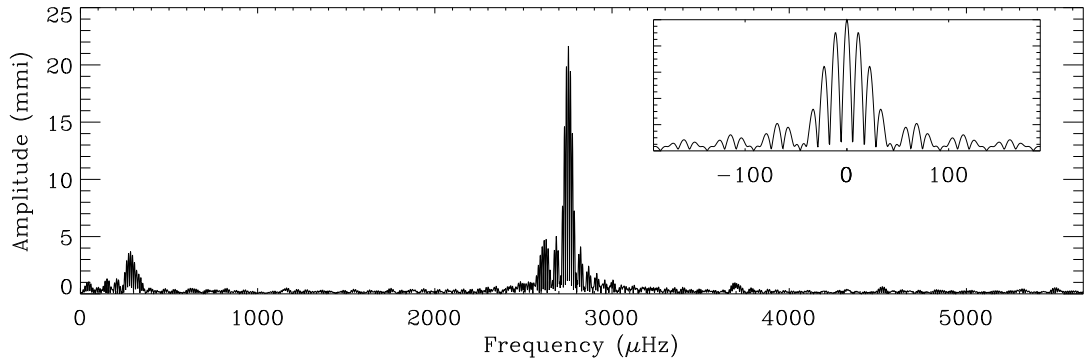}
  \caption{Discrete Fourier transform of the full light
  curve, with the window function for the two-night data set displayed
  in the inset panel in the upper right corner (frequency in
  $\mu$Hz as in the main plot, the amplitude scaling is
  arbitrary). $f_1$ and $f_2$ clearly stand out slightly below
  3000\,$\mu$Hz, as does the unresolved $f_3$ conglomeration near
  300\,$\mu$Hz. Note the ''shoulder'' of that feature towards higher
  frequencies which is responsible for significant residual power
  after prewhitening of one low-frequency period.}
  \label{fig:ft}
\end{figure*}
\begin{table}
  \begin{center}
    \caption[]{Frequencies extracted from the February 2004 data.}
    \label{tab:periods}
    \begin{tabular}{lrr@{}lr}
      \hline
      &frequency [$\mu$Hz]&period [s&]&amplitude [mmi]\\
      \hline
      $f_1$&2754\,(9) &  363.&1 & 21.7(5)\\
      $f_2$&2606\,(9) &  383.&7 &  4.6(5)\\
      $f_3$& 283\,(9) & 3538.&  &  3.7(5)\\
      \hline
    \end{tabular}
  \end{center}
\end{table}
Photometric observations of \object{HS\,0702$+$6043} were obtained on
three consecutive nights in December 1999 using the Calar Alto 1.2\,m
telescope. Details of the set-up, the observing log and the data
reduction may be found in \nocite{2002A&A...386..249D} DR02. 
The light curve has an average cycle
time of 18\,s and was taken in white light. The analysis of the
relative light curve obtained from aperture photometry yielded two
periods in the range of 360\,--\,390\,s as detailed in the previous
section.
\par
While testing a new set of time-series analysis tools (Huber et al.,
in preparation), we discovered a $\sim$1\,h variation in the residuals of
the discovery light curve of \object{HS\,0702$+$6043}. These data are however of
insufficient quality to conduct a false alarm probability analysis. In
order to confirm the variation, we reobserved the star during two
nights using the Steward Observatory 2.3\,m Bok telescope.
\par
These observations, obtained in February 2004, span roughly 6\,h on
both nights. A relative light curve was extracted from images that
have an average cycle time of 86\,s and were taken through a F555W
filter.
As in DR02, the two shorter periods (see Table~\ref{tab:periods}) were
prewhitened from the data. Note that the value DR02
\nocite{2002A&A...386..249D} have identified for their tentative
second short period is separated by a one-day alias
from the more reliable result for $f_2$ from the new data. 
\par
Both the full light curve and the residual after subtraction of $f_1$
and $f_2$ are shown in Fig.\,\ref{fig:lc}.
The frequency spectrum of the full data is displayed in Fig.\,\ref{fig:ft}. A
false alarm probability analysis of the residual (see DR02
\nocite{2002A&A...386..249D} for details
of our implementation of this method) for 
the peak near 271\,$\mu$Hz in Fig.\,\ref{fig:ft} indicates that it is
real with a significance well above 5$\sigma$. Moreover, there is
significant residual power left around this peak after prewhitening of
$f_3$. Regardless of ambiguities in the prewhitening due to one- and
two-day aliases, this is already evident in the original frequency spectrum
as a ''shoulder'' or ''hump'' to the right of the main peak. Since
this sub-structure is not resolved from what we have identified as
$f_3$, we do not however attempt to assign any further frequency
detections to it.
\section{What is causing the variation?}
\label{sec:discussion}
There are three possible interpretations of the low-frequency
variation: binarity (reflection effect or ellipsoidal variation), a
rotating star spot, or pulsation. Its coherence and persistence over
four years rules out normal sky or transparency variations.
\subsection{The binary hypothesis}
\label{sub:binary}
From the uncontaminated optical spectrum of \object{HS\,0702$+$6043},
we can easily exclude a potential F- or
G-type main sequence companion. Because the star is relatively faint,
\emph{2MASS} photometry is only accurate enough to require that any
hypothetical\linebreak companion must be later than K-type. 
At the supposed 1\,h orbit, the Roche geometry marginally allows to fit
an object near the minimum of the mass-radius relation, i.e.\ an
old massive brown dwarf.
However, all systems containing an sdB plus low-mass companion
show a substantial reflection effect. The amplitude of the observed variation
is quite low when compared with known sdB plus M-dwarf close double stars. \object{HW~Vir} was
the first such system discovered, with peak-to-peak amplitudes of
$\sim$0.26\,mag in $V$ \citep[e.g.][]{2000A&A...364..199K}, as well as
eclipses, while the most recent discovery is that of \object{HS\,2333$+$3927}
\citep{2004A&A...420..251H}, a system with
peak-to-peak amplitudes of 0.33\,mag in $R$. Another system, \object{PG\,1017$-$086},
shows a lower amplitude of 0.083 and a period of 1.75\,h
\citep{2002MNRAS.333..231M}, 
however this variation is still $\sim$22 times larger than the one we see in
\object{HS\,0702$+$6043}. One possible way to explain a very low-amplitude reflection is
by assuming a low inclination angle, i.e.\ so that the system is virtually
seen pole on; this is also the only way to prevent eclipses in very
tight configurations. 
\par
If the 1\,h variation we see in \object{HS\,0702$+$6043} is due to an ellipsoidal
distortion of the star by a compact companion, then we would expect that the
actual orbital period of the system is twice that of the intensity
variation. There are two known sdB plus white dwarf systems showing an
ellipsoidal effect, \object{KPD\,0422$+$5421} \citep{1998MNRAS.300..695K} and \object{KPD\,1930$+$2752}
\citep{2000ApJ...530..441B,2000MNRAS.317L..41M}.
These stars show intensity variations of 0.0148\,mag and
0.013\,mag, respectively, and both are seen almost edge on (with inclinations
$\ga$80$^\circ$). 
\object{Feige~48}, an \object{EC~14026} star with parameters very similar to
\object{HS\,0702$+$6043}, was found to be a sdB plus white dwarf
system of low inclination with a period of 9\,h \citep{2004A&A...422.1053O}.
If such a system were reduced to a 2\,h orbital period, ellipsoidal
variations might become apparent.
In addition to the inclination, the size of the variation depends on how much
of its Roche lobe the sdB fills. These relaxed constraints on the
amplitude together with a larger possible orbit (due to the twice as
long orbital period) to accommodate a second component that is more
compact make it harder to rule out the  sdB plus white dwarf scenario.
\par
However, if we examine the light curve after the two higher
pulsational frequencies have been subtracted (the thick line in
Fig.\,\ref{fig:lc}), it becomes quite evident that the amplitude of
the variation is changing. This strongly suggests that the variation
is not due to binary motion. 
\subsection{A rotating star spot?}
\label{sub:starspot}
Another possible cause to be considered as the origin of the variation
in \object{HS\,0702$+$6043} is a rotating star spot. Although
single sdBs do not generally rotate very rapidly, with only
one possible exception (\object{PG\,1605$+$072}, but see \citealt{2005astro.ph..6646K}),
the following scenario can be constructed around this
possibility. Assuming that the 1\,h period corresponds to the
rotational period of the star, the variation would then be caused by
an inhomogeneous surface temperature distribution (possibly resulting
from a magnetic field). The variation has been seen at the same period
in two observations taken four years apart, a time scale that is
probably much longer than the typical life time of a star
spot. However, newly appearing star spots are bound to result in the
same rotation period determination, only with arbitrary phase
discontinuities. Furthermore, differential rotation could explain the
variation in the period (''multiperiodicity'') when a single spot drifts,
or when several spots are present simultaneously.
\par
For the frequencies in the $p$-mode domain, substantial rotational
splitting due to the fast rotation would have to be taken into account
for all $l\ne0$ modes. For a hypothetical $l=1$ mode $f_0$ at a
suitable geometric orientation, the two $m=\pm1$ frequencies could be
observable and would lie at about $f_1=f_0+\Delta f$ and $f_2=f_0-\Delta f$,
where $\Delta f$ would correspond to the rotational frequency. Using
$f_1$ and $f_2$ from Table~\ref{tab:periods} and $f_0=(f_1+f_2)/2$,
$\Delta f$ would amount to 74\,$\mu$Hz, only a fraction of $f_3$.
\par
The fact that no frequency splitting at the large value of
$f_3$ (or multiples of it) is seen does not automatically rule out this
interpretation, but, together with the unusually high rotation rate
implied, makes it relatively unlikely.
\subsection{The long-period pulsation hypothesis}
\label{sub:pulsation}
The final possible cause of the observed variation is
pulsation. \citet{2003ApJ...583L..31G} first detected oscillations in
several sdBs with periods from around 30 to 80 minutes.
The periodicity we observe lies within this range, and the amplitude
is also very comparable to that of the \mbox{lpsdBVs}. 
As noted in the previous sections, the data imply multiperiodicity, a
strong indicator for a pulsational origin of the variation. This might
already be seen in the light curve, especially in the top panel of
Fig.\,\ref{fig:lc}, where the amplitude in the second half of the
curve appears to reduce almost to zero -- suggesting beating between
several modes. The unresolved multiperiodicity becomes particularly
evident after the highest peak in the low-frequency domain ($f_3$) has
been removed: After this additional prewhitening of the
strong variation, there is significant residual power remaining, above
the 3\,$\sigma$ level derived from a false alarm probability analysis.
\par
It is also interesting that the low frequency we measure
is very close to, but not exactly, the difference between the two
higher frequencies presented in Table~\ref{tab:periods}. This
near-resonance is striking, and again indicates that the variation is
most probably not due to rotation or orbital motion. If either of
these were the case, we would expect exact resonance.
All of this together suggests the very exciting possibility that
\object{HS\,0702$+$6043} shows both $p$-modes \emph{and} $g$-modes
simultaneously.  
\section{Implications for asteroseismology}
\label{sec:implications}
\begin{figure}
  \centering
  \includegraphics[width=0.4\textwidth]{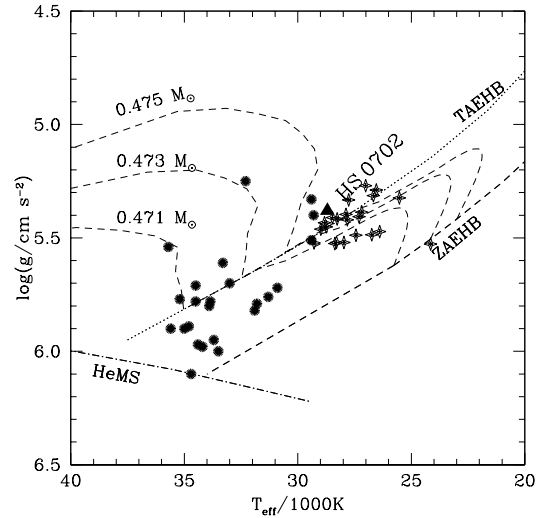}
  \caption{Known sdBV and lpsdBV pulsators 
    in the $\log{g}-{\rm T_{\rm eff}}$
    diagram: sdBV populate the higher-temperature, lpsdBV
    the lower-temperature region. The helium main sequence (HeMS),
    zero (ZAEHB) and terminal (TAEHB) age extended horizontal branches are
    marked; also shown are evolutionary tracks off the extended
    horizontal branch by \citet{dorman:93}. The position of
    \object{HS\,0702$+$6043} is indicated.}
  \label{fig:hrd}
\end{figure}

Stars that show observable multimode pulsations in both $p$-
\emph{and} $g$-modes have an enormous potential for asteroseismology
since modes on the acoustic and gravity branches, respectively, probe
different regions in the depth-dependent stellar
structure. \object{HS\,0702$+$6043} is probably one among very few
objects to fall into this class of hybrids. Not only does the range of its long
period(s) and the magnitude of the corresponding amplitude match what is
expected for $g$-mode pulsators, its stellar parameters also place
it at a position at the edge of both of the empirical instability
regions in the T$_{\rm eff} - \log{g}$ diagram (see Fig.\,\ref{fig:hrd}).
In fact, from its position in Fig.\,\ref{fig:hrd}, which shows the two
groups of hot, short-period pulsators, and the cooler, long-period
pulsators, one could argue that $g$-mode pulsations should almost be
\textit{expected} in \object{HS\,0702$+$6043} (although it should not
be forgotten that both types of pulsators among sdB stars co-exist
with a larger fraction of non-pulsating sdBs in the same areas).
\par
A second candidate for the class of \mbox{sdBV}\,/\,\mbox{lpsdBV}
hybrids established through the initial announcement of our discovery
\citep{2005ewwd.conf....2S} has recently been published by
\citet{2005MNRAS.360..737B}: \object{Balloon~090100001}
\citep{2004A&A...418..243O}.
It is striking that there appears to be a gap along the EHB,
so that the four coolest sdBVs (\object{Feige~48}, \object{HS\,2201$+$2610}, 
\object{Balloon~090100001} and \object{HS\,0702$+$6043}) are closely
connected in parameter space to the lpsdBVs rather than to the rest of the
sdBVs. 
In fact, two of them are hybrid (\object{HS\,0702$+$6043} and
\object{Balloon~090100001}), so what about \object{Feige~48} and
\object{HS\,2201$+$2610}?
If attempts to detect long-period variations remain unsucessful, the
latter two stars would belong to the sdBs in the lpsdBV regions that
do not show $g$-mode pulsations.
From a theoretical point of view, it is a challenge to drive both $p$-
and $g$-modes simultaneously in this object; future models will
have to account for the existence of \object{HS\,0702$+$6043}. 
\par
On the main sequence, where the \mbox{sdBV}\,/\,\mbox{lpsdBV} groups might
have an analogy in the $\beta$\,Cep\,/\,[SPB] variables, potential
hybrid stars include two objects presented by
\citet{1993AcA....43...13J} and \citet{2004MNRAS.347..454H}. 
The class of $\gamma$\,Dor\,/\,$\delta$\,Scu hybrids are discussed in
\citet{2002MNRAS.333..262H}, \citet{2005A&A...435..927D}, and
\citet{2005AJ....129.2026H}.
\section{The future}
\label{sec:outlook}
The amplitude variability observed in \object{HS\,0702$+$6043} is
highly suggestive that the 1\,h variation is not due to binary or
rotational motion, and results from $g$-mode pulsations instead. 
\par
To resolve the suspected several low-frequency
modes of \object{HS\,0702$+$6043}, longer observations are clearly
required. To this end, we have recently completed a 10-day-multisite campaign
to observe \object{HS\,0702$+$6043} photometrically that we will
report on in a forthcoming paper.
%

\begin{thebibliography}{23}
\expandafter\ifx\csname natexlab\endcsname\relax\def\natexlab#1{#1}\fi

\bibitem[{{Baran} {et~al.}(2005){Baran}, {Pigulski}, {Kozie{\l}}, {Og{\l}oza},
  {Silvotti}, \& {Zo{\l}a}}]{2005MNRAS.360..737B}
{Baran}, A., {Pigulski}, A., {Kozie{\l}}, D., {et~al.} 2005, \mnras, 360, 737

\bibitem[{{Bill{\`e}res} {et~al.}(2000){Bill{\`e}res}, {Fontaine}, {Brassard},
  {Charpinet}, {Liebert}, \& {Saffer}}]{2000ApJ...530..441B}
{Bill{\`e}res}, M., {Fontaine}, G., {Brassard}, P., {et~al.} 2000, \apj, 530,
  441

\bibitem[{{Charpinet} {et~al.}(2005){Charpinet}, {Fontaine}, {Brassard},
  {Chayer}, \& {Green}}]{charp_lapalma:05}
{Charpinet}, S., {Fontaine}, G., {Brassard}, P., {Chayer}, P., \& {Green}, E.
  2005, in Baltic Astronomy, Second Meeting on Hot Subdwarf Stars and Related
  Objects, in press

\bibitem[{{Charpinet} {et~al.}(1996){Charpinet}, {Fontaine}, {Brassard}, \&
  {Dorman}}]{charpinet:96}
{Charpinet}, S., {Fontaine}, G., {Brassard}, P., \& {Dorman}, B. 1996, ApJL,
  471, L103

\bibitem[{{Dorman} {et~al.}(1993){Dorman}, {Rood}, \& {O'Connell}}]{dorman:93}
{Dorman}, B., {Rood}, R.~T., \& {O'Connell}, R.~W. 1993, ApJ, 419, 596

\bibitem[{{Dreizler} {et~al.}(2002){Dreizler}, {Schuh}, {Deetjen}, {Edelmann},
  \& {Heber}}]{2002A&A...386..249D}
{Dreizler}, S., {Schuh}, S., {Deetjen}, J.~L., {Edelmann}, H., \& {Heber}, U.
  2002, A\&A, 386, 249

\bibitem[{{Dupret} {et~al.}(2005){Dupret}, {Grigahc{\` e}ne}, {Garrido},
  {Gabriel}, \& {Scuflaire}}]{2005A&A...435..927D}
{Dupret}, M.-A., {Grigahc{\` e}ne}, A., {Garrido}, R., {Gabriel}, M., \&
  {Scuflaire}, R. 2005, \aap, 435, 927

\bibitem[{{Fontaine} {et~al.}(2003){Fontaine}, {Brassard}, {Charpinet},
  {Green}, {Chayer}, {Bill{\`e}res}, \& {Randall}}]{2003ApJ...597..518F}
{Fontaine}, G., {Brassard}, P., {Charpinet}, S., {et~al.} 2003, \apj, 597, 518

\bibitem[{{Green} {et~al.}(2003){Green}, {Fontaine}, {Reed}, {Callerame},
  {Seitenzahl}, {White}, {Hyde}, {{\O}stensen}, {Cordes}, {Brassard}, {Falter},
  {Jeffery}, {Dreizler}, {Schuh}, {Giovanni}, {Edelmann}, {Rigby}, \&
  {Bronowska}}]{2003ApJ...583L..31G}
{Green}, E.~M., {Fontaine}, G., {Reed}, M.~D., {et~al.} 2003, \apjl, 583, L31

\bibitem[{{Han} {et~al.}(2003){Han}, {Podsiadlowski}, {Maxted}, \&
  {Marsh}}]{2003MNRAS.341..669H}
{Han}, Z., {Podsiadlowski}, P., {Maxted}, P.~F.~L., \& {Marsh}, T.~R. 2003,
  \mnras, 341, 669

\bibitem[{{Handler} {et~al.}(2002){Handler}, {Balona}, {Shobbrook}, {Koen},
  {Bruch}, {Romero-Colmenero}, {Pamyatnykh}, {Willems}, {Eyer}, {James}, \&
  {Maas}}]{2002MNRAS.333..262H}
{Handler}, G., {Balona}, L.~A., {Shobbrook}, R.~R., {et~al.} 2002, \mnras, 333,
  262

\bibitem[{{Handler} {et~al.}(2004){Handler}, {Shobbrook}, {Jerzykiewicz},
  {Krisciunas}, {Tshenye}, {Rodr{\'{\i}}guez}, {Costa}, {Zhou}, {Medupe},
  {Phorah}, {Garrido}, {Amado}, {Papar{\' o}}, {Zsuffa}, {Ramokgali}, {Crowe},
  {Purves}, {Avila}, {Knight}, {Brassfield}, {Kilmartin}, \&
  {Cottrell}}]{2004MNRAS.347..454H}
{Handler}, G., {Shobbrook}, R.~R., {Jerzykiewicz}, M., {et~al.} 2004, \mnras,
  347, 454

\bibitem[{{Heber} {et~al.}(2004){Heber}, {Drechsel}, {{\O}stensen}, {Karl},
  {Napiwotzki}, {Altmann}, {Cordes}, {Solheim}, {Voss}, {Koester}, \&
  {Folkes}}]{2004A&A...420..251H}
{Heber}, U., {Drechsel}, H., {{\O}stensen}, R., {et~al.} 2004, \aap, 420, 251

\bibitem[{{Henry} \& {Fekel}(2005)}]{2005AJ....129.2026H}
{Henry}, G.~W. \& {Fekel}, F.~C. 2005, \aj, 129, 2026

\bibitem[{{Jerzykiewicz}(1993)}]{1993AcA....43...13J}
{Jerzykiewicz}, M. 1993, Acta Astronomica, 43, 13

\bibitem[{{Kiss} {et~al.}(2000){Kiss}, {Cs{\' a}k}, {Szatm{\' a}ry}, {Fur{\'
  e}sz}, \& {Szil{\' a}di}}]{2000A&A...364..199K}
{Kiss}, L.~L., {Cs{\' a}k}, B., {Szatm{\' a}ry}, K., {Fur{\' e}sz}, G., \&
  {Szil{\' a}di}, K. 2000, \aap, 364, 199

\bibitem[{{Koen} {et~al.}(1998){Koen}, {Orosz}, \&
  {Wade}}]{1998MNRAS.300..695K}
{Koen}, C., {Orosz}, J.~A., \& {Wade}, R.~A. 1998, \mnras, 300, 695

\bibitem[{{Kuassivi} {et~al.}(2005){Kuassivi}, {Bonanno}, \&
  {Ferlet}}]{2005astro.ph..6646K}
{Kuassivi}, {Bonanno}, A., \& {Ferlet}, R. 2005, astro-ph/0506646

\bibitem[{{Maxted} {et~al.}(2002){Maxted}, {Marsh}, {Heber}, {Morales-Rueda},
  {North}, \& {Lawson}}]{2002MNRAS.333..231M}
{Maxted}, P.~F.~L., {Marsh}, T.~R., {Heber}, U., {et~al.} 2002, \mnras, 333,
  231

\bibitem[{{Maxted} {et~al.}(2000){Maxted}, {Marsh}, \&
  {North}}]{2000MNRAS.317L..41M}
{Maxted}, P.~F.~L., {Marsh}, T.~R., \& {North}, R.~C. 2000, \mnras, 317, L41

\bibitem[{{Oreiro} {et~al.}(2004){Oreiro}, {Ulla}, {P{\' e}rez Hern{\' a}ndez},
  {{\O}stensen}, {Rodr{\'{\i}}guez L{\' o}pez}, \&
  {MacDonald}}]{2004A&A...418..243O}
{Oreiro}, R., {Ulla}, A., {P{\' e}rez Hern{\' a}ndez}, F., {et~al.} 2004, \aap,
  418, 243

\bibitem[{{O'Toole} {et~al.}(2004){O'Toole}, {Heber}, \&
  {Benjamin}}]{2004A&A...422.1053O}
{O'Toole}, S.~J., {Heber}, U., \& {Benjamin}, R.~A. 2004, \aap, 422, 1053

\bibitem[{{Schuh} {et~al.}(2005){Schuh}, {Huber}, {Green}, {O'Toole},
  {Dreizler}, {Heber}, \& {Fontaine}}]{2005ewwd.conf....2S}
{Schuh}, S., {Huber}, J., {Green}, E.~M., {et~al.} 2005, in ASP Conf. Series
  334: 14$^{\rm th}$ European Workshop on White Dwarfs (2004), ed. D.~{Koester}
  \& S.~{M{\"o}hler}, 530

\end{thebibliography}

%
%
\end{document}